\documentclass[aps,prc,twocolumn,groupedaddress]{revtex4}

\usepackage{graphicx}
\begin{document}

%\preprint{}

\title{On the Surface Structure of Strange Superheavy Nuclei}


\author{Jeff McIntire}
\email{jwmcin@wm.edu}
%\thanks{}
\affiliation{Deparment of Physics, College of William and Mary,
Williamsburg, VA 23187}


\date{\today}

\begin{abstract}
Bound, strange, neutral superheavy nuclei, stable 
against strong decay, may exist. A model effective 
field theory calculation of the surface energy and 
density of such systems is carried out assuming 
vector meson couplings to conserved currents and 
scalar couplings fit to data where it exists. The
non-linear relativistic mean field equations are
solved assuming local density baryon sources. The 
approach is calibrated through a successful 
calculation of the known nuclear surface tension.
\end{abstract}

%\pacs{}
%\keywords{}

\maketitle


\section{Introduction}
An inspection of the semiempirical mass formula 
(SEMF) reveals that the single largest limiting 
factor in the creation of very large nuclei is 
the Coulomb repulsion. One way to overcome this barrier 
is to include hyperons in nuclei \cite{ref:Sc93,ref:Sc00,ref:St99}. 
Consider the hyperons $\Lambda^{0}$, $\Sigma^{(\pm,0)}$, and $\Xi^{(-,0)}$.
The lightest hyperon, the $\Lambda$, has a negative
binding energy in nuclear matter \cite{ref:Ga88} 
and decays weakly into nonstrange matter. The $\Sigma$'s appear to 
have a repulsive nuclear potential \cite{ref:Ga95,ref:Ba99,ref:Da99}.
Next in mass are the $\Xi$'s; experimental evidence suggests
that the binding energy of a single $\Xi$ in 
nuclear matter is negative \cite{ref:Ga01,ref:Fu98,ref:Kh00}. 
In addition, the reaction $2 \Lambda \rightarrow N + 
\Xi $ becomes energetically favorable for some critical 
number of $\Lambda$'s in the nuclear 
medium \cite{ref:Sc93}. As a result,
we expect that for large systems the addition of $\Lambda$'s
and $\Xi$'s is desirable, but the inclusion of $\Sigma$'s 
would have little or no positive effect. Therefore we 
consider matter composed solely of p, n, $\Lambda$, 
$\Xi^{0}$, and $\Xi^{-}$. The inclusion 
of the $\Xi^{-}$'s offsets the Coulomb repulsion 
of the protons; this potentially allows for the creation 
of arbitrarily large nuclei by diminishing the importance
of the Coulomb term in the SEMF. To minimize the effect
of the Coulomb term we investigate this class of nuclei
such that $\mathrm{Q} = 0$. Due to the fact that these nuclei are stable 
against strong decay, they consequently decay on weak 
interaction timescales enhancing the potential for their detection. 
The purpose of this paper is to model the surface structure, and acquire 
the surface energy from the calculated SEMF, for this 
class of nuclei.

To accomplish this, we must solve the nuclear 
many-body problem. Using quantum hadrodynamics (QHD), we 
construct an effective lagrangian density invariant under $SU(2)_{L} 
\bigotimes SU(2)_{R}$ symmetry using hadrons 
as degrees of freedom \cite{ref:Fu96,ref:Wa95}. Isodoublet baryon fields, 
$\psi$, are coupled to the meson fields $\phi$, V, $\rho$, and $\pi$;
the meson fields correspond to an isoscalar, Lorentz scalar 
and vector, and an isovector, vector and pseudoscalar, fields 
respectively \cite{ref:Se97}. In the relativistic
mean field limit, the meson fields become classical 
fields and the sources become expectation
values. The pions have no mean fields in a spherically symmetric 
system. Then the lagrangian density is converted into 
a hamiltonian density; since it is explicitly time independent 
the hamiltonian density is equivalent to the energy density. 
The baryon mass is determined by solving the scalar field 
equation self-consistently at each point \cite{ref:Wa95,ref:Se95}.
The coupling constants are fit to reproduce experimental 
values of various ordinary nuclei; specifically the 
parameter sets NLC and Q1, which include nonlinear 
scalar couplings, are used \cite{ref:Se97,ref:Fu96}. This 
theory, when solved in the relativistic Hartree approximation,
has had great success in predicting the bulk 
properties and shell structure in ordinary nuclei
\cite{ref:Fu90,ref:Fu96}. 

To investigate the structure of the many-particle superheavy
nuclei of interest here, we directly use the approach of density
functional theory (DFT). DFT tells us that the 
exact ground-state density is acquired by minimizing
the energy functional \cite{ref:Se97,ref:Ko99}. The effective
lagrangian discussed above provides a lowest order density 
functional. The scalar and vector fields here play the role
of relativistic Kohn-Sham potentials \cite{ref:Ko99,ref:Se00}.
In order to calibrate our approach, we calculate 
the ground-state densities of ordinary finite nuclei with 
N = Z. To model finite nuclei we must add spherically 
symmetric spatial variations of the meson fields to the 
lagrangian density. The source terms
are evaluated using a local density approximation; at 
every point within the nucleus the baryons are assumed 
to be a local Fermi gas with states filled up to $k_{F}$. We 
acquire the scalar mean field equation by
minimizing the energy functional with respect to the scalar
field; a similar approach yields the vector mean field equation
\cite{ref:Se95}. 

The nonlinear scalar field equation is solved as a finite difference
equation utilizing a shooting method. The boundary conditions are
determined by assuming the baryon density vanishes at the surface
and solving the linear scalar field equation outside
\cite{ref:Se95}. However, these boundary conditions are 
exact only in the linear case; a correction term must 
be added to compensate for the effects of nonlinear terms in 
the scalar field equation. The total energy is minimized with 
respect to the local Fermi wave number, while keeping fixed the 
baryon number, B. The constraint of fixed B is 
incorporated with a Lagrange multiplier, which is 
the chemical potential. The resulting constraint 
equation states that the chemical potential must be constant 
throughout the nucleus \cite{ref:Se95}. This approach is more 
sophisticated than a simple Thomas-Fermi method because
we self-consistently solve for the source terms at each point.
With the calculated binding energy and baryon number for finite
nuclei, we fit the first two terms in the SEMF, the bulk and 
surface energies, for nuclear matter. These are in
good agreement with the known experimental values \cite{ref:Wa95},
thereby validating our approach. 

Now we add in hyperons; however, existing experimental 
data requires that some assumptions be made. 
First, we couple universally to the conserved 
baryon and isovector currents. Second, a different 
scalar coupling is used for each baryon. 
The scalar coupling for the $\Lambda$'s is fit
such that the binding energy of a single $\Lambda$
in nuclear matter is -28 MeV \cite{ref:Ga88}. However the 
binding energy of a single $\Xi$ is relatively uncertain, 
values appearing in the literature range from
-40 to -14 MeV. Recent experiments with light nuclei suggest
that the value lies on the less bound side of this 
range \cite{ref:Fu98,ref:Kh00};
however, it may be more deeply bound for heavy nuclei \cite{ref:Ya94}.
As a result, a number of values for the
$\Xi$ scalar coupling are investigated. Third, we continue
to utilize the parameter sets for ordinary nuclear matter, NLC
and Q1, to generate the nucleon and non-linear scalar couplings.

The addition of new baryons in the theory only requires 
the inclusion of new source terms in the effective 
lagrangian density. We investigate a specific sector of 
the theory by imposing the restrictions $\mathrm{Q} = 0$ 
and $\mathrm{|S| / B} = 1$, where S is the total strangeness. 
We also assume an average cascade mass. Note that the minimum
binding energy always occurs such that there are equal numbers of
n and p (and consequently equal numbers of $\Xi^{0}$ and $\Xi^{-}$); 
therefore the symmetry term
in the SEMF is rendered irrelevant. Since there is only
one chemical potential, the reactions $2 \Lambda \leftrightarrow N + \Xi$ 
and $n + \Xi^{0} \leftrightarrow p + \Xi^{-}$ are in
equilibrium. Again, using DFT to 
model finite nuclei, we now investigate the role of 
the first two terms in the calculated SEMF. Also, by determining
the baryon density, we acquire the structure of the surface.

At $T = 0$ and normal nuclear densities, the mass difference 
between strange and nonstrange quarks is less then the 
Fermi energy of massless nonstrange quarks. This opened the
possibility that strange quark matter composed of u, d, 
and s quarks might be stable against strong decay
and perhaps even absolutely stable \cite{ref:Wi84,ref:Ja84}.
These systems are characterized by small charge 
fraction $\mathrm{Q / B} \sim 0$ and large strangeness fraction 
$\mathrm{|S| / B} \sim 1$. The plausibility of bound strange 
matter has also been explored in the hadronic sector. Theoretical
investigations of multiple $\Lambda$ hypernuclei indicate 
that they are bound and stable against strong decay. These 
studies produced systems with binding energies as low as 
-9 MeV corresponding to $\mathrm{|S| / B} \sim 0.2$ 
\cite{ref:Ma93,ref:Ru90,ref:Ba91,ref:La92,ref:Sc98}.
 
Gal et.\ al.\ suggested that discussions of matter composed 
of n, p, and $\Lambda$ must also include $\Xi^{0}$ and $\Xi^{-}$ 
due to the fact that the reaction $2 \Lambda \rightarrow N + \Xi$ 
is energetically favorable for some critical number of $\Lambda$'s 
in the nuclear medium \cite{ref:Sc93}. They investigated possible 
configurations of bulk matter in the relativistic mean field 
approach, suggesting binding energies per 
baryon as low as -25 MeV with a large strangeness fraction; finite
nuclear calculations were also preformed  
\cite{ref:Sc00,ref:Ga95,ref:Do95,ref:Gr94}. In addition, Gal 
et.\ al.\ fit to a generalized SEMF using a Fermi gas model 
\cite{ref:Do93,ref:Ba94}. Extrapolating
from the ordinary SEMF they estimate the bulk and symmetry terms,
while leaving the Coulomb term unchanged. The surface energy is scaled as 
inversely proportional to the average baryon mass, yielding a value
of 15 MeV. Stoks and Lee challenged 
these findings using a many-body theory with baryon-baryon 
potential models. These potentials were developed using 
an $SU(3)$ extension of the Nijmegen soft-core potentials 
\cite{ref:Ya99,ref:Ri99,ref:Ha99,ref:Ri98}. In contrast, 
the latter found that this type of matter is only slightly bound, 
$\mathrm{E / B} \sim -3$ MeV or less \cite{ref:St99,ref:Vi98}. A 
quark-meson coupling model produced a minimum binding energy 
of -24.4 MeV with $\mathrm{|S| / B} \sim 1.38$ \cite{ref:Wo99}. The
effect of adding hyperons has also been explored in application to
neutron stars \cite{ref:Ga97,ref:Ba00,ref:Bu00,ref:Gl97}.

In this paper we first construct the effective
lagrangian of the theory following the methodology of
Furnstahl, Serot, and Tang \cite{ref:Se97,ref:Fu96,ref:Fu97}. 
For ordinary nuclear matter, we model both infinite matter and
finite nuclei of various radii using QHD.
Initially, it was assumed that, because of the relatively large
value of vector meson mass, the derivatives of the 
vector field were negligible \cite{ref:Se95}; however, 
iteration on the vector field produces a significant 
effect on the calculation. Also, a correction to the 
boundary condition is required to account for the nonlinearities in the 
scalar field equation. From this calculation, we acquire
the baryon density and scalar field for ordinary finite nuclei; this 
provides a picture of the size and shape of the surface. 
Then the surface energy is extracted by fitting to the SEMF. 
The approach is successfully calibrated by comparing this result to the 
experimentally obtained values. Next this system is extended 
to cascade-nucleon matter, composed of p, n, $\Xi^{0}$, and 
$\Xi^{-}$, subject to the constraints $\mathrm{Q} = 0$ and 
$\mathrm{|S| / B} = 1$. However, due to the uncertainty in the 
binding of a single cascade in nuclear matter, we examine a range 
of values for the $\Xi$ scalar coupling. Again, by calculating 
finite systems, the bulk and surface terms are 
extracted from the SEMF. $\Lambda$'s are then
added to the cascade-nucleon matter, with the $\Lambda$ scalar
coupling fit to experiment. The bulk
quantities change little, however, from the cascade-nucleon
case. An investigation of possible hyperon-hyperon interaction
is conducted by coupling a $\Phi$ meson to the conserved 
strangeness current; we allow the $\Phi$ coupling to increase 
until the many-body system is no longer bound in order to find 
the maximum allowable value of this coupling.

The organization of this paper is as follows. In 
section 2, the effective lagrangian is constructed and
the systems of equations are derived. In section 3, we 
present the methodologies used to solve these systems
described in section 2. In section 4, the results for 
nuclear and hyperon-nucleon matter are discussed.

\section{Theory}
Following the QHD approach of Furnstahl et.\ al.\ 
\cite{ref:Fu96,ref:Fu97,ref:Se97}, we construct an effective 
lagrangian density using hadronic degrees
of freedom that remains invariant under 
$SU(2)_{L} \bigotimes SU(2)_{R}$ symmetry. We will use this
lagrangian density to model both infinite and finite systems;
to start with only systems of nucleons with $N = Z$, which we
refer to as \emph{nucleon} matter, are considered. The full lagrangian 
density (with $\hbar = c = 1$) is given by
\begin{eqnarray}
{\cal{L}}(x_{\mu}) & = & -\bar{\psi}\left[\gamma_{\mu}\left(\frac{\partial}
{\partial x_{\mu}} - ig_{v}V_{\mu}\right) + M - g_{s}\phi\right]\psi
\nonumber \\ & & - \frac{1}{2}\left[\left(\frac{\partial \phi}
{\partial x_{\mu}}\right)^{2} + m_{s}^{2}\phi^{2}\right]
- \frac{1}{4}F_{\mu\nu}F_{\mu\nu} \nonumber \\ & &
- \frac{1}{2}m_{v}^{2}V_{\mu}^{2}
- \frac{\kappa}{3!}\phi^{3} - \frac{\lambda}{4!}\phi^{4}
\label{eq:LA1}
\end{eqnarray}

\noindent Here $\psi = \left( \begin{array}{c} p \\ n 
\end{array} \right)$ is the isodoublet baryon field, 
$\gamma_{\mu} = (i\vec{\alpha}\beta,\beta)$ are the gamma matrices, 
and $\mu,\nu = 1,\ldots4$. The conventions of \cite{ref:Wa95} are used.
The Lorentz scalar meson field, $\phi$, is coupled to the
scalar density $\bar{\psi}\psi$ and $V_{\mu}$, the Lorentz 
vector meson field, is coupled to the conserved baryon current
$i\bar{\psi}\gamma_{\mu}\psi$;
their respective coupling constants are $g_{s}$ and $g_{v}$.
The masses of the nucleon, scalar, and vector fields are 
denoted by $M$, $m_{s}$, and $m_{v}$ respectively. $\kappa$ 
and $\lambda$ are constants that determine the strength of 
the nonlinear scalar couplings. The field tensor is defined as
\begin{equation}
F_{\mu\nu} = \frac{\partial V_{\nu}}{\partial x_{\mu}} 
- \frac{\partial V_{\mu}}{\partial x_{\nu}}
\label{eq:fieldT}
\end{equation}

\noindent 
Notice the $\rho$-meson terms in Eq.\ (\ref{eq:LA1}) have 
been suppressed. The source term contributed by the 
$\rho$-meson depends on $N - Z$, which vanishes 
for the systems under consideration. We now employ 
relativistic mean field theory (RMFT). In this case 
the meson fields are replaced by their classical, or 
mean fields
\begin{eqnarray}
\phi(x_{\mu}) & \longrightarrow & \phi_{0}(r) \\
V_{\mu}(x_{\mu}) & \longrightarrow & i\delta_{\mu4}V_{0}(r)
\end{eqnarray}

\noindent which are time-independent. For the purposes of this
paper, we restrict ourselves to spherical symmetry. The source 
terms are now replaced 
by their expectation values. In DFT, the classical fields play
the role of Kohn-Sham potentials \cite{ref:Se97,ref:Se00}. Incorporating 
both Eq.\ (\ref{eq:fieldT}) and RMFT, our lagrangian density becomes
\begin{eqnarray}
{\cal{L}}(x_{\mu}) & = & -\bar{\psi}\left(\gamma_{\mu}\frac{\partial}
{\partial x_{\mu}} + M^{*}\right)\psi - g_{v}V_{0}\psi^{\dagger}\psi
\nonumber \\ & & + \frac{1}{2}[(\nabla V_{0})^{2} 
+ m_{v}^{2}V_{0}^{2}] - \frac{1}{2}[(\nabla \phi_{0})^{2} 
+ m_{s}^{2}\phi_{0}^{2}] \nonumber \\ & &
- \frac{\kappa}{3!}\phi_{0}^{3} - \frac{\lambda}{4!}\phi_{0}^{4}
\end{eqnarray}

\noindent where the effective mass is defined as 
\begin{equation}
M^{*} \equiv M - g_{s}\phi_{0}
\end{equation}

\noindent Now the hamiltonian density is given by
\begin{eqnarray}
{\cal{H}}(x_{\mu}) & = & \Pi\frac{\partial \psi}{\partial t} - {\cal{L}}
\nonumber \\ & 
= & \psi^{\dagger}\frac{\partial \psi}{\partial t}
+ \bar{\psi}\left(\gamma_{\mu}\frac{\partial}{\partial x_{\mu}} 
+ M^{*}\right)\psi + g_{v}\rho_{B}V_{0} \nonumber \\ & & 
- \frac{1}{2}[(\nabla V_{0})^{2} + m_{v}^{2}V_{0}^{2}] 
+ \frac{1}{2}[(\nabla \phi_{0})^{2} + m_{s}^{2}\phi_{0}^{2}]
\nonumber \\ & & 
+ \frac{\kappa}{3!}\phi_{0}^{3} + \frac{\lambda}{4!}\phi_{0}^{4}
\end{eqnarray}

\noindent where $\rho_{B} = \psi^{\dagger}\psi$ is the baryon 
density and the canonical momentum density is
\begin{equation}
\Pi = \frac{\partial {\cal{L}}}{\partial (\partial \psi /
\partial t)} = i\psi^{\dagger}
\end{equation}

\noindent If one assumes that a nucleus is a local Fermi gas filled 
up to some $k_{F}(r)$ at every point, the source terms take the form
\begin{equation}
\rho_{B}(r) = \langle\psi^{\dagger}\psi\rangle = \frac{\gamma}{(2\pi)^{3}}
\int_{0}^{k_{F}(r)} d^{3}k
\end{equation}
\begin{equation}
\langle\psi^{\dagger}(\vec{\alpha}\cdot\vec{p} + \beta M^{*})\psi\rangle =
\frac{\gamma}{(2\pi)^{3}}\int_{0}^{k_{F}(r)}(k^{2} + M^{*2})^{1/2}d^{3}k
\end{equation}

\noindent where $\vec{p} = - i\nabla$ and $\gamma$
is a degeneracy factor. Since the hamiltonian density is 
explicitly time-independent, it is equivalent to the energy 
density (with $\gamma = 4$ for nucleon matter)
\begin{eqnarray}
{\cal{E}}(r) & = & \frac{1}{2}[(\nabla\phi_{0})^{2} + m_{s}^{2}\phi_{0}^{2}] 
- \frac{1}{2}[(\nabla V_{0})^{2} + m_{v}^{2}V_{0}^{2}] \nonumber \\ & &
+ g_{v}\rho_{B}V_{0} + \frac{\kappa}{3!}\phi_{0}^{3} 
+ \frac{\lambda}{4!}\phi_{0}^{4} \nonumber \\ & &  
+ \frac{4}{(2\pi)^{3}}\int_{0}^{k_{F}}(k^{2} + M^{*2})^{1/2}d^{3}k
\label{eq:ED}
\end{eqnarray}

\noindent The total energy and baryon number are
\begin{equation}
E = \int{\cal{E}}(r)d^{3}r; \:\;\;\; B = \int\rho_{B}(r)d^{3}r 
\end{equation}

\noindent The energy density above provides a lowest order 
density functional. DFT tells us that minimizing the exact 
energy functional yields the exact ground-state density.

To conduct infinite nucleon matter calculations, we neglect 
spatial variations in the meson fields; the resulting 
energy density is 
\begin{eqnarray}
{\cal{E}} & = & \frac{1}{2}m_{s}^{2}\phi_{0}^{2} 
+ \frac{\kappa}{3!}\phi_{0}^{3} + \frac{\lambda}{4!}\phi_{0}^{4}
+ g_{v}\rho_{B}V_{0} - \frac{1}{2} m_{v}^{2}V_{0}^{2}
\nonumber \\ & &
+ \frac{4}{(2\pi)^{3}}\int_{0}^{k_{F}}(k^{2} + M^{*2})^{1/2}d^{3}k
\label{eq:IE}
\end{eqnarray}

\noindent By minimizing the energy functional with respect 
to the scalar field, the scalar mean field equation is determined. 
The vector mean field equation is similarly derived as
an extremum of the energy functional. These equations are
\begin{eqnarray}
\phi_{0} + \frac{\kappa}{2m_{s}^{2}}\phi_{0}^{2}
& + & \frac{\lambda}{6m_{s}^{2}}\phi_{0}^{3} = 
\frac{g_{s}}{m_{s}^{2}}\rho_{s} \label{eq:IS} \\
V_{0} & = & \frac{g_{v}}{m_{v}^{2}}\rho_{B}
\label{eq:IV}
\end{eqnarray}

\noindent where the scalar density is given by
\begin{equation}
\rho_{s} = \frac{4}{(2\pi)^{3}}\int_{0}^{k_{F}} 
d^{3}k\frac{M^{*}}{(k^{2} + M^{*2})^{1/2}}
\end{equation} 

\noindent The solution to these equations is discussed in
section 3.

We now turn our attention to finite nucleon systems.  
We retain the spherically symmetric spatial variations in the 
meson fields and therefore require the full energy density
in Eq.\ (\ref{eq:ED}). The meson field equations, acquired in the 
same manner as above, are
\begin{eqnarray}
(\nabla^{2} - m_{s}^{2})\phi_{0} - \frac{\kappa}{2}\phi_{0}^{2}
& - & \frac{\lambda}{6}\phi_{0}^{3} = - g_{s}\rho_{s}
\label{eq:SE1} \\
(\nabla^{2} - m_{v}^{2})V_{0} & = & - g_{v}\rho_{B}
\label{eq:VE1}
\end{eqnarray}

\noindent Note that for spherically symmetric systems, the 
laplacian becomes
\begin{equation}
\nabla^{2} = \frac{\partial^{2}}{\partial r^{2}} + \frac{2}{r}
\frac{\partial}{\partial r}
\end{equation}

\noindent Using a Green's function, the solution to Eq.\ (\ref{eq:VE1}) is 
\begin{eqnarray}
g_{v}V_{0} & = & \frac{g_{v}^{2}}{4\pi}\int d^{3}y\rho_{B}(\vec{y})
\frac{e^{-m_{v}|\vec{x} - \vec{y}|}}{|\vec{x} - \vec{y}|} \\ 
& = & \frac{g_{v}^{2}}{xm_{v}}\int ydy\rho_{B}(y) 
\sinh (m_{v}x_{\scriptscriptstyle <}) 
e^{-m_{v}x_{\scriptscriptstyle >}}
\label{eq:GF}
\end{eqnarray}

\noindent where the angular dependence has now been 
integrated out. Since the contribution of the laplacian in 
Eq.\ (\ref{eq:VE1}) is small compared with that of the 
vector meson mass, to a first approximation it can be 
neglected \cite{ref:Se95,ref:Wa95},
\begin{equation}
g_{v}V_{0} = \frac{g_{v}^{2}}{m_{v}^{2}}\rho_{B}
\label{eq:VEA}
\end{equation}

\noindent However, the omitted term produces a small, but important, 
contribution to the vector field; this can have a significant 
effect on the total energy and baryon number. It is then convenient 
to express the vector field as
\begin{equation}
g_{v}V_{0} = \frac{g_{v}^{2}}{m_{v}^{2}}\rho_{B} + \delta W_{0}
\label{eq:VE2}
\end{equation}

\noindent with $\delta W_{0} \equiv g_{v}\delta V_{0}$. 
Substituting this in Eq.\ (\ref{eq:GF}) and rearranging, 
one obtains an explicit expression for $\delta W_{0}$ in 
terms of the baryon density
\begin{equation}
\delta W_{0}(r) = \frac{g_{v}^{2}}{4\pi}\int d^{3}y\rho_{B}(\vec{y})
\frac{e^{-m_{v}|\vec{x} - \vec{y}|}}{|\vec{x} - \vec{y}|}
- \frac{g_{v}^{2}}{m_{v}^{2}}\rho_{B}(r)
\label{eq:dV}
\end{equation}

Minimization of the total energy with respect 
to the local Fermi wave number now yields the ground 
state of the system. A Lagrange multiplier
is used to incorporate the constraint of fixed B such that 
\begin{equation}
\delta E(k_{F},\phi_{0},V_{0}) - \mu\delta B(k_{F}) = 0
\end{equation}

\noindent Since the variations of the energy density with 
respect to both the scalar and vector 
field vanish, they can be held constant in the 
variation of $k_{F}$, the result of which is the constraint
equation 
\begin{equation}
\mu = g_{v}V_{0}(r) + [k_{F}^{2}(r) + M^{*2}(r)]^{1/2}
\end{equation}

\noindent where the Lagrange multiplier, $\mu$, is the 
chemical potential and is constant throughout
the nucleus. 

On the surface $r = r_{0}$, the baryon density vanishes. 
The constraint equation at the surface then yields the 
first boundary condition
\begin{equation}
M^{*}(r_{0}) = \mu - \delta W_{0}(r_{0})
\label{eq:BC1}
\end{equation}

\noindent where Eq.\ (\ref{eq:VE2}) has now been employed. 

To determine the second boundary condition, consider the 
solution to the linear homogeneous scalar field equation
\begin{equation}
\phi_{0} = \phi_{c}\frac{e^{-(r-r_{0})m_{s}}}{r/r_{0}}
\label{eq:SES}
\end{equation}

\noindent The constant $g_{s}\phi_{c} = M - \mu + \delta W_{0}(r_{0})$ is 
determined at the surface. Differentiating Eq.\ (\ref{eq:SES}) with 
respect to $r$ and then evaluating at the surface, we acquire the 
second boundary condition 
\begin{equation}
\left[\frac{\partial M^{*}(r)}{\partial r}\right]_{r_{0}}  = 
[M + \delta W_{0}(r_{0}) - \mu]\left(\frac{1 + m_{s}r_{0}}{r_{0}}\right)
(1 + \epsilon)
\label{eq:BC2}
\end{equation}

\noindent The solution to the scalar field equation in Eq.\ 
(\ref{eq:SES}) no longer holds when the nonlinear terms 
are included; therefore, a small correction $\epsilon$ has been 
included to compensate. In the nonlinear case, the scalar field
equation is integrated outward from  $r_{0}$, and $\epsilon$ 
in Eq.\ (\ref{eq:BC2}) is varied until the solution 
vanishes for large $r$.

The calculated binding energy and baryon number 
of a series of nuclei with different radii can be fit with a
SEMF of the form
\begin{equation}
\frac{E}{B} = a_{1} + a_{2}B^{-1/3}
\label{eq:SEMF}
\end{equation}

\noindent where only the bulk and surface terms have been retained. 
The bulk constant, $a_{1}$, is determined by the binding
energy of infinite nuclear matter. Then, after calculating a number 
of finite nuclei, the surface energy, $a_{2}$, can be obtained
by plotting the calculated energies per baryon against the calculated 
values of $B^{-1/3}$. The results of numerical methods for these
finite systems are discussed in later sections.

We now extend our theory to consider systems of nucleons 
and hyperons. First we investigate matter composed of n, p, 
$\Xi^{0}$, and $\Xi^{-}$, subject to the conditions 
\begin{eqnarray}
& Q = 0 &
\label{eq:Q0} \\
& |S| / B = 1 &
\label{eq:S/B}
\end{eqnarray}

\noindent where $Q$ and $S$ are the total charge and strangeness
respectively. These systems shall be subsequently referred to as 
\emph{cascade-nucleon} ($\Xi N$) matter. Equation (\ref{eq:Q0}) 
restricts the system to equal numbers of p and $\Xi^{-}$; 
similarly, Eq.\ (\ref{eq:S/B}) forces the numbers of n and 
$\Xi^{-}$ to be equal. Therefore the system is now characterized 
by two Fermi wave numbers, $k_{Fp}$ and $k_{Fn}$. For simplicity 
we employ an average cascade mass $M_{\Xi} = (M_{\Xi^{0}} + 
M_{\Xi^{-}}) / 2$. Since the energy density is now symmetric under 
the interchange of $k_{Fp}$ and $k_{Fn}$, the minimum binding 
energy always occurs such that $k_{Fp} = k_{Fn}$. As a result, we 
can further restrict this system to a single Fermi wave number, $k_{F}$. 
It is a consequence of these arguments that equilibrium is imposed 
upon the reaction 
\begin{equation}
n + \Xi^{0} \rightleftharpoons p + \Xi^{-}
\label{eq:RE1}
\end{equation}

\noindent and the system is described by only one chemical 
potential. Again we mention that the $\rho$-mesons do not 
contribute here for similar reasons to the \emph{nucleon} case. 

Next, we must make some assumptions about the cascade couplings.
Since the baryon current is conserved, the vector coupling 
is taken to be universal, $g_{v} = g_{v\Xi}$. However, an 
independent scalar coupling for the cascades, $g_{s\Xi}$, is 
assumed.

Consider the case of infinite $\Xi N$ matter. The addition of 
hyperons to the theory requires only the addition of new 
source terms. A source term of the form
\begin{equation}
\delta{\cal{E}} = \frac{\gamma}{(2\pi)^{3}}\int_{0}^{k_{F}}(k^{2} 
+ M_{\Xi}^{*2})^{1/2}d^{3}k
\label{eq:dE2}
\end{equation}

\noindent is added to the energy density in Eq.\ (\ref{eq:IE})
where
\begin{equation}
M_{\Xi}^{*} = M_{\Xi} - g_{s\Xi}\phi_{0}
\end{equation}

\noindent In addition, a new term is included in the baryon density
\begin{equation}
\delta \rho_{B} = \frac{\gamma}{(2\pi)^{3}}\int_{0}^{k_{F}}d^{3}k
\label{eq:dpb2}
\end{equation}

\noindent Here $\gamma = 4$ for ($\Xi^{0}$,$\Xi^{-}$) with spin
up and down. Except for the additional source terms, the meson field 
equations remain unchanged. The new term added to the source in
the scalar field Eq.\ (\ref{eq:IS}) is
\begin{equation}
\delta \rho_{s} = \frac{4 s}{(2\pi)^{3}}\int_{0}^{k_{F}} 
d^{3}k\frac{M_{\Xi}^{*}}{(k^{2} + M_{\Xi}^{*2})^{1/2}}
\label{eq:dps2}
\end{equation}

\noindent where
\begin{equation}
s \equiv g_{s\Xi} / g_{s}
\end{equation}

\noindent Equation (\ref{eq:dpb2}) is incorporated into 
the source in the vector field Eq.\ (\ref{eq:IV}). The solution
to these equations is again discussed in section 3.

We now examine the case of finite $\Xi N$ matter.
The source terms in Eqs.\ (\ref{eq:dE2}) and (\ref{eq:dpb2}) are 
incorporated into the energy density in Eq.\ (\ref{eq:ED}). 
Next the terms in Eqs.\ (\ref{eq:dps2}) and (\ref{eq:dpb2}) 
are added to meson field equations in Eqs.\ (\ref{eq:SE1}) 
and (\ref{eq:VE1}) respectively. Then a new constraint equation 
is produced in the same manner as before
\begin{equation}
\mu = g_{v}V_{0} + \frac{1}{2}[(k_{F}^{2} + M^{*2})^{1/2}
+ (k_{F}^{2} + M_{\Xi}^{*2})^{1/2}]
\end{equation}

\noindent Similarly, the boundary conditions are now
\begin{equation}
M^{*}(r_{0}) = \frac{1}{1 + s}[2\mu - 2\delta W_{0}(r_{0})
- M_{\Xi} + sM]
\end{equation}
\begin{eqnarray}
\left[\frac{\partial M^{*}(r)}{\partial r}\right]_{r_{0}} & = &
\frac{1}{1 + s}[2\delta W_{0}(r_{0}) - 2\mu + M_{\Xi} \nonumber \\ & & 
+ M]\left(\frac{1 + m_{s}r_{0}}{r_{0}}\right)(1 + \epsilon)
\end{eqnarray}

\noindent Consider again the SEMF in Eq.\ (\ref{eq:SEMF}). The 
conditions imposed on $\Xi N$ matter in Eqs.\ (\ref{eq:Q0}) and
(\ref{eq:S/B}) now justify the elimination of the Coulomb and symmetry 
terms. Then $a_{1}$ is taken to be the binding energy of infinite
$\Xi N$ matter; next, proceeding as before, the surface energy,
$a_{2}$, can be extracted.

Finally, we investigate a class of matter in which 
$\Lambda$'s are added to the $\Xi N$ matter described 
above. These systems are referred to as \emph{lambda-cascade-nucleon} 
($\Lambda\Xi N$) matter. The previous restrictions do not 
relate the number of $\Lambda$'s to the number of $N$'s and 
$\Xi$'s; therefore a second Fermi wave 
number, $k_{F\Lambda}$, is needed. Again the vector coupling is 
taken to be universal and an independent scalar coupling, 
$g_{s\Lambda}$, is employed. Now equilibrium is imposed on 
the reactions
\begin{eqnarray}
n + \Xi^{0} & \rightleftharpoons & \Lambda + \Lambda \\
p + \Xi^{-} & \rightleftharpoons & \Lambda + \Lambda
\end{eqnarray}

\noindent as well as on Eq.\ (\ref{eq:RE1}). As before, the system is 
then characterized by a single chemical potential.
The source terms required for the inclusion of $\Lambda$'s in
both infinite and finite $\Xi N$ matter are
\begin{equation}
\delta{\cal{E}} = \frac{2}{(2\pi)^{3}}\int_{0}^{k_{F\Lambda}}
(k^{2} + M_{\Lambda}^{*2})^{1/2}d^{3}k
\end{equation}
\begin{equation}
\delta \rho_{B} = \frac{2}{(2\pi)^{3}}\int_{0}^{k_{F\Lambda}}d^{3}k
\end{equation}
\begin{equation}
\delta \rho_{s} = \frac{2 t}{(2\pi)^{3}}\int_{0}^{k_{F\Lambda}} 
d^{3}k\frac{M_{\Lambda}^{*}}{(k^{2} + M_{\Lambda}^{*2})^{1/2}}
\end{equation}

\noindent where 
\begin{equation}
M^{*}_{\Lambda} = M_{\Lambda} - g_{s\Lambda}\phi_{0}
\end{equation}

\noindent and 
\begin{equation}
t \equiv g_{s\Lambda} / g_{s}
\end{equation}

\noindent In the case of finite $\Lambda\Xi N$ matter, 
there are now two constraint equations
\begin{equation}
\mu = g_{v}V_{0} + \frac{1}{2}[(k_{F}^{2} + M^{*2})^{1/2}
+ (k_{F}^{2} + M_{\Xi}^{*2})^{1/2}]
\end{equation}
\begin{equation}
\mu = g_{v}V_{0} + (k_{F\Lambda}^{2} + M_{\Lambda}^{*2})^{1/2}
\end{equation}

\noindent The $\Lambda$ density begins interior to the surface
$r_{0}$; this allows the $\Xi N$ boundary conditions to be used
in the $\Lambda\Xi N$ case.

\begin{figure}
\includegraphics[width=\columnwidth]{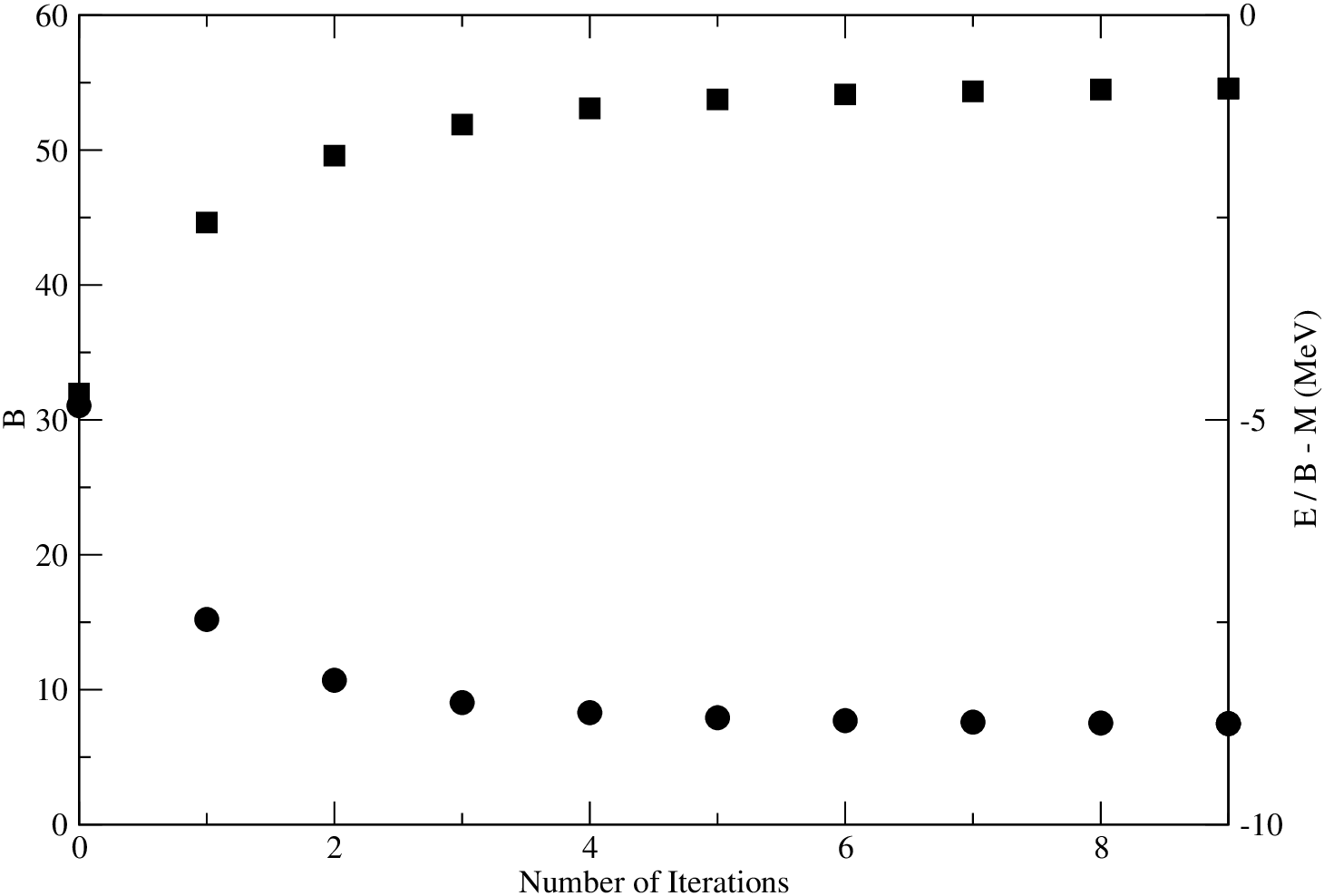}
\caption{Convergence of the baryon number and binding energy 
per baryon (boxes and circles respectively) after 9 iterations
for an ordinary nucleus of $r_{0} = 15 / m_{s}$, N = Z, and using
the L2 parameter set.}
\label{fig:conver}
\end{figure}

\begin{table}
\begin{tabular}{|c|c|c|c|c|c|c|c|} \hline
  &$g_{s}^{2}$  &$g_{v}^{2}$  &$m_{s}$  &$m_{v}$  &$\kappa$  
&$\lambda$ & $g_{s\Lambda} / g_{s}$ \\ \hline
L2   &109.63  &190.43  &520    &783  &0      &0        &.886601  \\ \hline 
NLC  &95.11   &148.93  &500.8  &783  &5000   &-200     &.881898 \\ \hline
Q1   &103.67  &164.70  &504.57 &782  &4577.6 &-197.70  &.884029  \\ \hline
\end{tabular}
\caption{Parameter sets taken from \cite{ref:Fu97,ref:Se97}. $m_{s}$, 
$m_{v}$, $\kappa$, and $\lambda$ are in MeV. $g_{s\Lambda}
/ g_{s}$ is fit to reproduce the binding energy
of a single $\Lambda$ in nuclear matter \cite{ref:Ga88}.}
\label{tab:PAR}
\end{table}

\begin{table}[t!]
\begin{tabular}{|c|c|} \hline
$g_{s\Xi} / g_{s}$ &$\delta\cal{E} / \delta\rho_{B} - 
\mathrm{M_{\Xi}}$ \\ \hline
1    &-68.9  \\ \hline 
.95  &-51.6  \\ \hline
.9   &-34.3  \\ \hline
\end{tabular}
\caption{Values of the binding energy (in MeV) of a single $\Xi$
in nuclear matter for various $\Xi$ coupling ratios and the
parameter set NLC.}
\label{tab:CCBE}
\end{table}

\section{Methodology}
In this section we develop a methodology for solving
the systems of equations discussed in section 2.  
In the case of nucleon matter the parameters 
$g_{s}$, $g_{v}$, $m_{s}$, $m_{v}$, $\mathrm{M}$, $\kappa$, 
and $\lambda$ must first be specified. The vector meson mass 
is defined to be the mass of the $\omega$-meson and $\mathrm{M} 
\equiv 939$ MeV. The 
remaining constants are given by the three coupling sets
in Table \ref{tab:PAR}; to determine these sets the
theory was fit to reproduce various properties of ordinary 
nuclear matter \cite{ref:Fu97,ref:Se97}. The simplest set, L2, 
includes only linear terms in the scalar field. The sets 
NLC and Q1 both expand the theory to include nonlinear terms. 
As a result, these sets must be fit to more properties
of nuclear matter than L2.

To extend the theory to systems of nucleons and hyperons, 
specification of the constants $g_{s\Lambda}$, $g_{s\Xi}$, 
$g_{v\Lambda}$, and $g_{v\Xi}$ is also required. Since the 
vector meson is coupled to the conserved baryon current, we 
assume a universal vector coupling, $g_{v} = g_{v\Lambda} = 
g_{v\Xi}$. The scalar couplings, on the other hand, are 
adjusted to reproduce the binding energies of single hyperons 
in nuclear matter. For instance, the $\Lambda$ scalar coupling 
is designed to reproduce the binding energy of a single 
$\Lambda$ in nuclear matter, experimentally determined to 
be $-28$ MeV \cite{ref:Ga88}. The values of $g_{s\Lambda} / g_{s}$ 
are also shown in Table \ref{tab:PAR}. Unfortunately data 
on the binding energy of a single $\Xi$ in nuclear matter 
is uncertain. Therefore, a range of $\Xi$ scalar couplings 
is investigated; the values used are $g_{s\Xi} / g_{s} = 1$, $.95$, 
and $.9$. These values correspond to the binding energies 
listed in Table \ref{tab:CCBE}. 

Consider the case of infinite nucleon matter. To obtain the 
solution to this system, first one must specify $k_{F}$. Both 
Eqs.\ (\ref{eq:IS}) and (\ref{eq:IV}) are now solved for their 
respective meson fields. Then, using the meson fields and $k_{F}$, 
one calculates the energy density in Eq.\ (\ref{eq:IE}). This is 
in turn used to evaluate the binding energy per baryon, $\mathrm{BE}
(k_{F}) \equiv {\cal{E}} / \rho_{B} - \mathrm{M}$. In RMFT the medium
saturates and $\mathrm{BE}(k_{F})$ has a minimum; 
this equilibrium value, $\mathrm{BE_{0}}$, serves as
the bulk term in the SEMF, $a_{1}$. This procedure is also 
applicable to infinite $\Xi N$ and $\Lambda\Xi N$ systems 
provided the appropriate source terms are included.

Now we discuss the methodology used for all finite systems. 
First the scalar field Eq.\ (\ref{eq:SE1}) is converted into a pair 
of coupled first-order finite difference equations for [$\phi_{0}(r)$,
$\phi_{0}'(r)$]; these equations are solved using a shooting method. 
To accomplish this, one fixes $\mu$ and $r_{0}$; now the boundary 
conditions are uniquely determined. Since $k_{F}(r_{0}) = 0$, 
we can solve the finite difference equations for 
$\phi_{0}(r)$ and $\phi_{0}'(r)$ one step in. These
solutions are substituted back into the constraint equation from
which $k_{F}(r)$, one step in, is determined. In
this manner we iterate in from the surface, evaluating $\rho_{B}(r)$ 
and ${\cal{E}}(r)$ at every point. This process is repeated until
$\rho_{B}$ and $\phi_{0}$ become constant as $r$ approaches the 
origin (or equivalently $\rho_{B}' = \phi_{0}' = 0$ at $r = 0$); 
we achieve this by adjusting the chemical potential. We mention that
better convergence is obtained by decreasing the step size.
Note that initially, $\epsilon$ and $\delta W_{0}$ are ignored.

Now we incorporate the two correction terms. In order to discuss the
role of $\epsilon$, let us examine the second boundary condition. 
When nonlinear terms are 
introduced, the small parameter, $\epsilon$, is included to 
compensate. Iterating the finite difference equations \emph{out} 
from the surface, $\epsilon$ in Eq.\ (\ref{eq:BC2}) 
is adjusted such that the scalar field vanishes for large $r$. 
The newly corrected boundary condition is then used to 
resolve the finite system by integrating \emph{in} as described above. 

\begin{figure}
\includegraphics[width=\columnwidth]{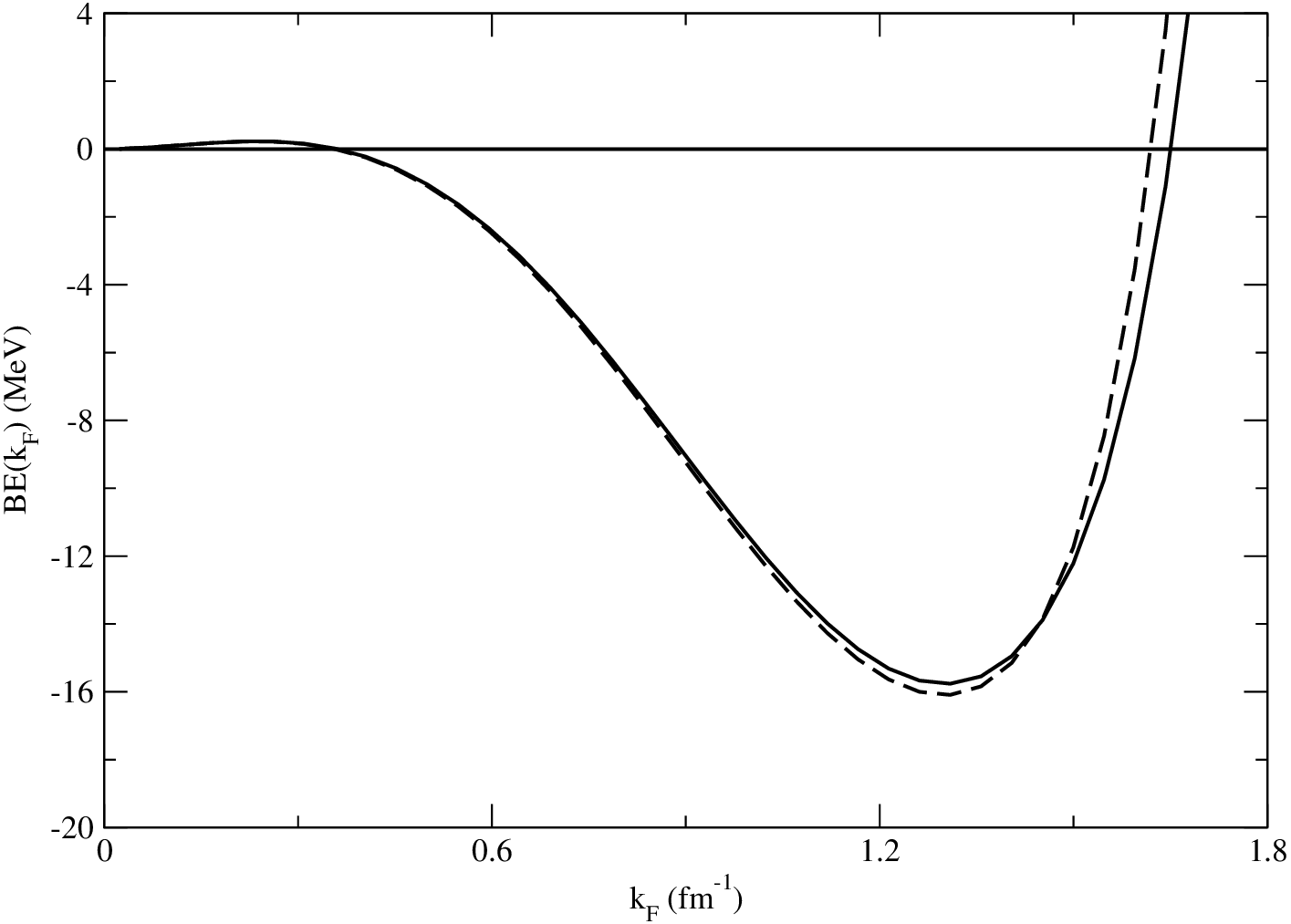}
\caption{Binding energy per baryon for infinite nuclear matter 
with N = Z as a function of Fermi wave number. These results 
are for the coupling sets NLC (solid line) and Q1 (dashed line).}
\label{fig:NLCQ1}
\end{figure}

\begin{table}
\begin{tabular}{|c|c|c|c|} \hline
      &$k_{F}$  &$\mathrm{M^{*} / M}$  
&$\cal{E} / \rho_{B} - \mathrm{M}$ \\ \hline
L2    &1.301    &.5409        &-15.758     \\ \hline 
NLC   &1.301    &.6313        &-15.768     \\ \hline
Q1    &1.299    &.5975        &-16.099     \\ \hline
\end{tabular}
\caption{Calculated equilibrium values of the Fermi momentum 
(in $\mathrm{fm^{-1}}$), effective mass, and the $\mathrm{BE_{0}}$ 
(in MeV) for infinite nucleon matter are shown using the 
coupling sets in Table \ref{tab:PAR}. These numbers 
reproduce the results in \cite{ref:Fu97,ref:Se97}.}
\label{tab:IN}
\end{table}

\begin{table}[t!]
\begin{tabular}{|c|c|c|c|c|} \hline
      &$g_{s\Xi} / g_{s}$  &$k_{F}$  &$\mathrm{M^{*} / M}$ 
&$\cal{E} / \rho_{B} - \mathrm{M_{\Lambda}}$  \\ \hline
NLC   &1   &1.363   &.1391  &-41.343  \\ \cline{2-5} 
      &.95 &1.351   &.1686  &-22.747  \\ \cline{2-5}
      &.9  &1.326   &.2232  &-4.741   \\ \hline
Q1    &1   &1.320   &.1372  &-42.728  \\ \cline{2-5}
      &.95 &1.310   &.1643  &-24.013  \\ \cline{2-5}
      &.9  &1.290   &.2104  &-5.812   \\ \hline
\end{tabular}
\caption{Calculated equilibrium values of the Fermi momentum 
(in $\mathrm{fm^{-1}}$), effective mass, and the $\mathrm{BE_{0}}$ 
(in MeV) for infinite $\Xi N$ matter are shown using 
the coupling sets NLC and Q1 in Table \ref{tab:PAR} and 
a range of values for $g_{s\Xi} / g_{s}$.}
\label{tab:ICASN}
\end{table}

Next, the correction term $\delta W_{0}$ is added to the 
vector field. Initially Eq.\ 
(\ref{eq:VEA}) was employed; however, this is accurate only in the 
limit of a large $m_{v}$. A small, but important, contribution 
to the vector field was omitted; therefore, the term $\delta W_{0}$,
defined by Eq.\ (\ref{eq:dV}) and calculated from the previous 
$\rho_{B}(r)$, is included. Then the entire process is repeated 
again. After successive iterations on the vector field, 
B and $\mathrm{BE} \equiv \mathrm{E} / \mathrm{B} - \mathrm{M}$
both converge to their full solutions. This 
convergence is illustrated in Fig.\ \ref{fig:conver} for 
ordinary nucleon matter; here, the parameter set L2 was used,
a radius of $r_{0} = 15m_{0}^{-1}$ was assumed, and 9 iterations
on the vector field have been carried out. Similar convergence
was found in all cases studied here.

Finally, we consider the SEMF in Eq.\ (\ref{eq:SEMF}) where
$a_{1}$ is defined as the $\mathrm{BE}_{0}$ of infinite matter. 
The calculated values of BE and $\mathrm{B}^{-1/3}$ of the 
finite systems are plotted against each other
for nuclei of various radii. Then using this SEMF as a linear
fit, the surface energy, $a_{2}$, is determined. The above 
approach to finite systems is first calibrated by the \emph{nucleon} 
matter case; then it is extended to investigate the $\Xi N$ and 
$\Lambda\Xi N$ systems detailed in section 2.

\section{Results and Discussion}
In this section, we discuss the application of the above
methodologies. First we consider the results of
our infinite matter calculations, starting with nucleon systems. 
Table \ref{tab:IN} shows the equilibrium values of $k_{F}$, 
$\mathrm{M^{*} / M}$, and $\mathrm{BE_{0}}$ obtained for 
infinite nucleon matter using the L2, NLC, and Q1 
parameter sets. The values calculated here reproduce 
those in \cite{ref:Fu97,ref:Se97}. Then $\mathrm{BE}(k_{F})$ is plotted 
for both the NLC and Q1 sets in Fig.\ \ref{fig:NLCQ1}. 
The minimum, $\mathrm{BE}_{0}$, in Fig.\ \ref{fig:NLCQ1} is 
taken to be $a_{1}$ in the SEMF for each coupling set. 
This is in good agreement with the empirical bulk term
\cite{ref:Wa95}.

\begin{figure}
\includegraphics[width=\columnwidth]{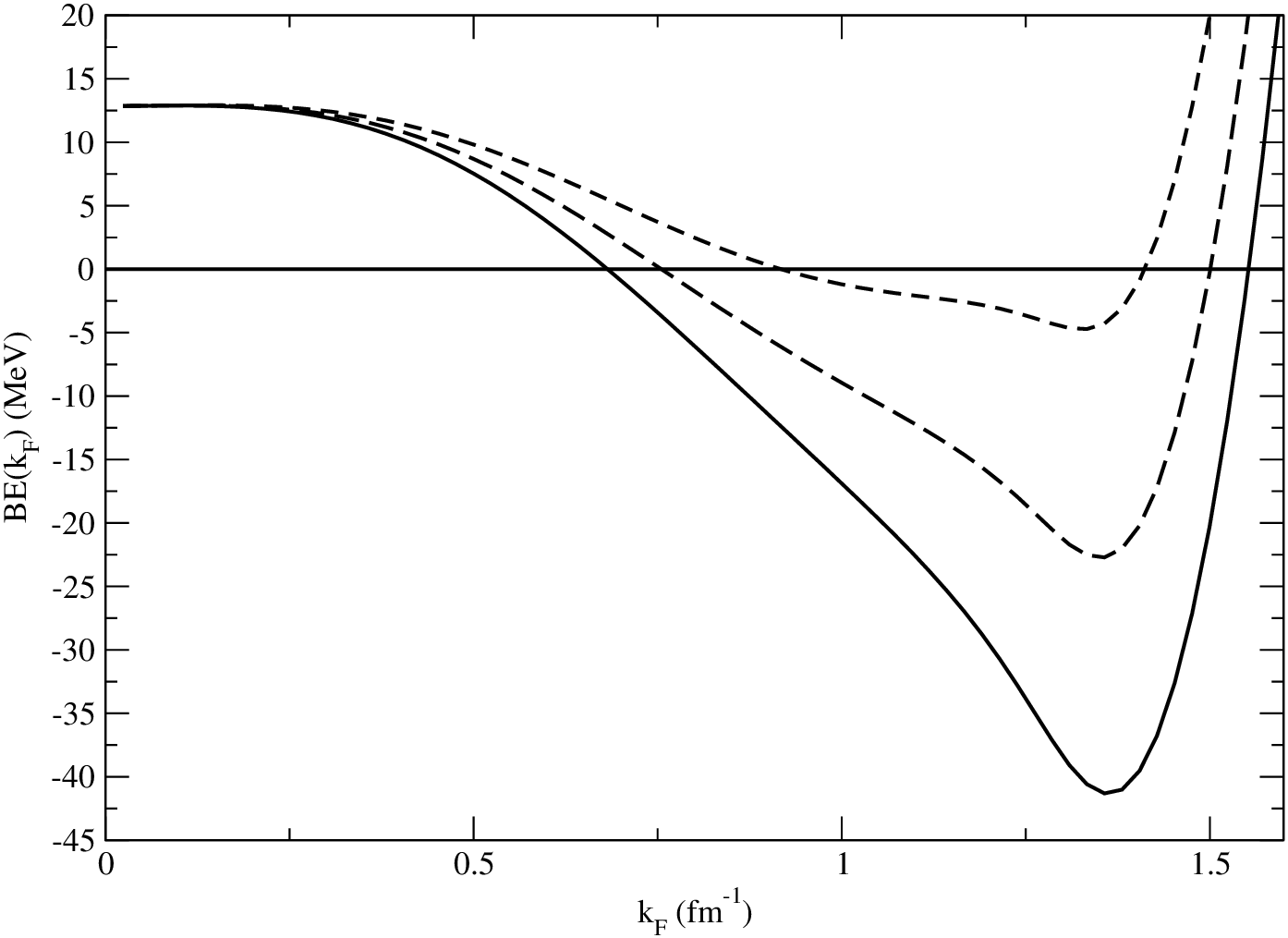}
\caption{Binding energies per baryon for infinite cascade-nucleon 
matter computed relative to isolated lambdas (the lowest energy 
free baryon state for $\mathrm{|S| / B} = 1$) as a function of the Fermi 
wave number using NLC. Note the left hand intercept is 
$\mathrm{(M_{\Xi} + M_{N}) / 2 - M_{\Lambda}}$. The solid, long dashed,
and short dashed lines correspond to $g_{s\Xi} / g_{s}$ = 1, 
0.95, and 0.9 respectively.}
\label{fig:res5NLC2b}
\end{figure}

We now turn our attention to infinite systems of nucleons 
and hyperons; these investigations are conducted using only 
the nonlinear cases, NLC and Q1. We begin by investigating infinite 
$\Xi N$ matter for the range of $\Xi$ scalar couplings 
mentioned above. Since the lowest mass state of separated 
baryons under the conditions Eqs.\ (\ref{eq:Q0}) and (\ref{eq:S/B})
consists entirely of $\Lambda$'s, $\mathrm{BE}(k_{F})$ 
for infinite cascade-nucleon matter is defined by 
$\mathrm{BE}(k_{F}) \equiv \cal{E} / \rho_{B} - \mathrm{M_{\Lambda}}$. 
The calculated equilibrium values for 
this type of matter are given in Table \ref{tab:ICASN}. 
Notice that $\mathrm{|BE_{0}|}$ decreases as the $\Xi$ 
coupling grows weaker while the equilibrium $k_{F}$
remains fairly constant. Graphs of $\mathrm{BE}(k_{F})$ for the 
NLC set and each $\Xi$ scalar coupling ratio 
shown in Fig.\ \ref{fig:res5NLC2b} illustrate this point. 
Although the equilibrium $k_{F}$ is 
roughly the same here as in the nucleon case, these systems contain 
twice as many baryons; as a result, the baryon density is much higher
than in infinite nucleon matter. Also, the effective 
mass is considerably smaller, on the order of a third the value 
of the nucleon case. Again $a_{1}$ in the SEMF
for each coupling ratio is taken to be the minimum 
($\mathrm{BE_{0}}$) of the corresponding curve in 
Fig.\ \ref{fig:res5NLC2b}. We also mention that the 
value of $g_{s\Xi} / g_{s} = .886$ is the lowest $\Xi$ 
coupling ratio for which infinite $\Xi N$ matter was 
still bound.

\begin{table}
\begin{tabular}{|c|c|c|c|c|c|c|} \hline
      &$g_{s\Xi} / g_{s}$  &$k_{F}$  &$k_{F\Lambda}$  &$\mathrm{M^{*} / M}$  
&$\cal{E} / \rho_{B} - \mathrm{M_{\Lambda}}$  \\ \hline
NLC   &1   &1.343   &.8665  &.1202  &-42.229  \\ \cline{2-6} 
      &.95 &1.319   &1.026  &.1321  &-24.704  \\ \cline{2-6}
      &.9  &1.288   &1.159  &.1511  &-8.100   \\ \hline
Q1    &1   &1.302   &.8141  &.1213  &-43.457  \\ \cline{2-6}
      &.95 &1.278   &.9831  &.1325  &-25.792  \\ \cline{2-6}
      &.9  &1.247   &1.124  &.1495  &-9.095   \\ \hline
\end{tabular}
\caption{Calculated equilibrium values of the Fermi momenta 
(in $\mathrm{fm^{-1}}$), effective mass, and the $\mathrm{BE_{0}}$ 
(in MeV) for infinite $\Lambda\Xi N$ matter are 
shown using the coupling sets NLC and Q1 in Table \ref{tab:PAR} 
and a range of values for $g_{s\Xi} / g_{s}$.}
\label{tab:ILCN}
\end{table}

Next, we consider infinite $\Lambda\Xi N$ systems using the same
range of $\Xi$ scalar couplings quoted above. Note that as in 
the $\Xi N$ case, $\mathrm{BE}(k_{F}) \equiv \cal{E} / \rho_{B} 
- \mathrm{M_{\Lambda}}$. This investigation produced the equilibrium 
values shown in Table \ref{tab:ILCN}. Here 
the equilibrium values of $k_{F}$, $\mathrm{M_{*} / M}$, and 
$\mathrm{BE_{0}}$ differ little from the $\Xi N$ results 
for large $\Xi$ coupling; however, the difference becomes more 
pronounced as the $\Xi$ coupling decreases. In our formulation,
a second Fermi wave number, $k_{F\Lambda}$, was included for 
the $\Lambda$'s; as one might expect, $k_{F\Lambda}$ grows,
and consequently the proportion of $\Lambda$'s increases,
as the gap between $g_{s\Xi}$ and $g_{s\Lambda}$ narrows. 
The smallest value of the $\Xi$ scalar 
coupling for which infinite $\Lambda\Xi N$ matter was 
still bound was $g_{s\Xi} / g_{s} = .875$.

\begin{figure}
\includegraphics[width=\columnwidth]{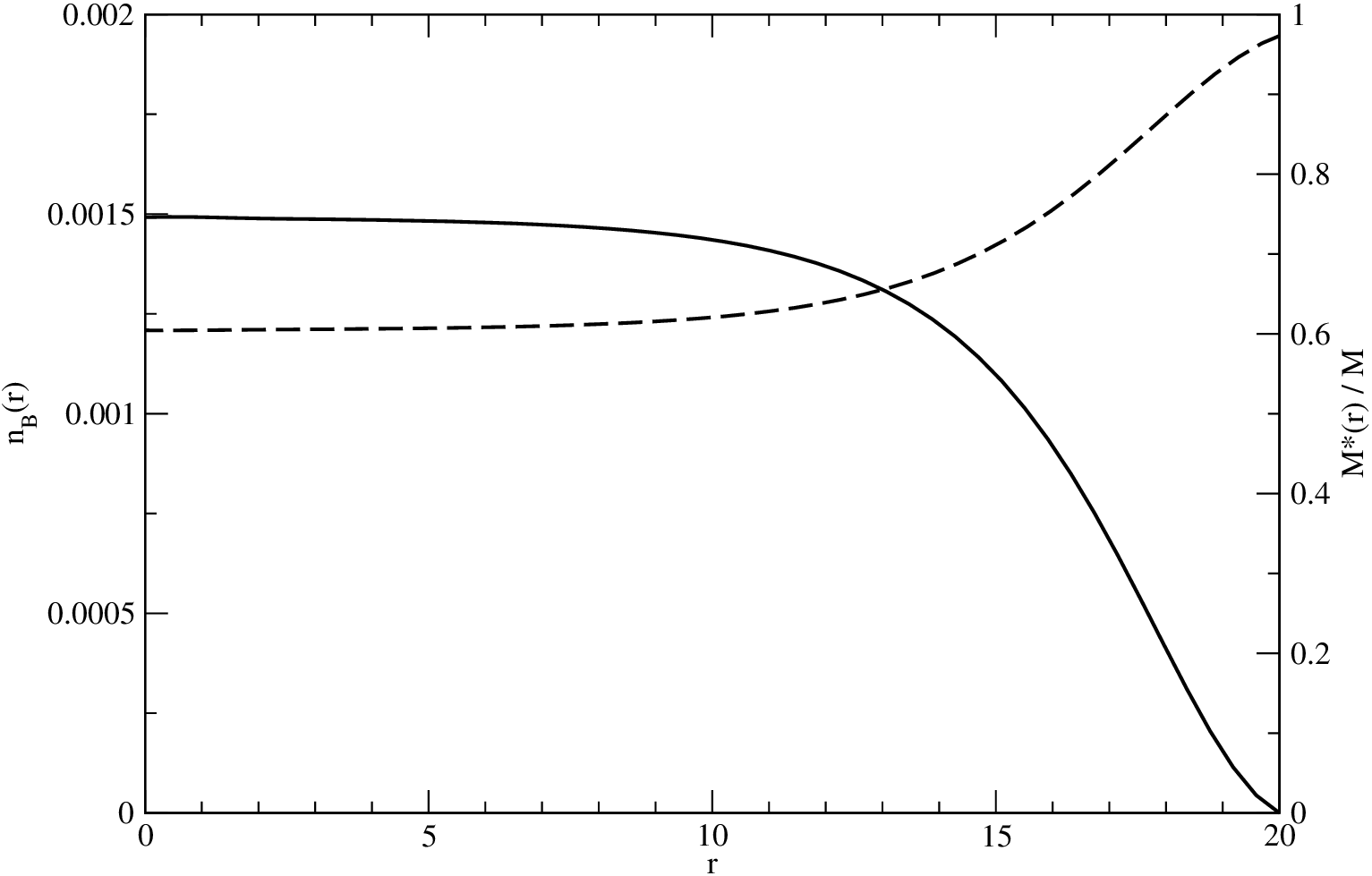}
\caption{The baryon density $n_{B}(r) = \rho_{B}(r) / \mathrm{M^{3}}$ 
(solid line) and effective mass $\mathrm{M^{*}(r) / M}$ (dashed line) 
vs. r (in units of $m_{s}^{-1}$) for an ordinary finite 
nucleus with N = Z, $\mathrm{B} = 188.869$ $r_{0} = 20 / m_{s}$, and 
using the NLC parameter set.}
\label{fig:tf21NLCb}
\end{figure}

\begin{figure}
\includegraphics[width=\columnwidth]{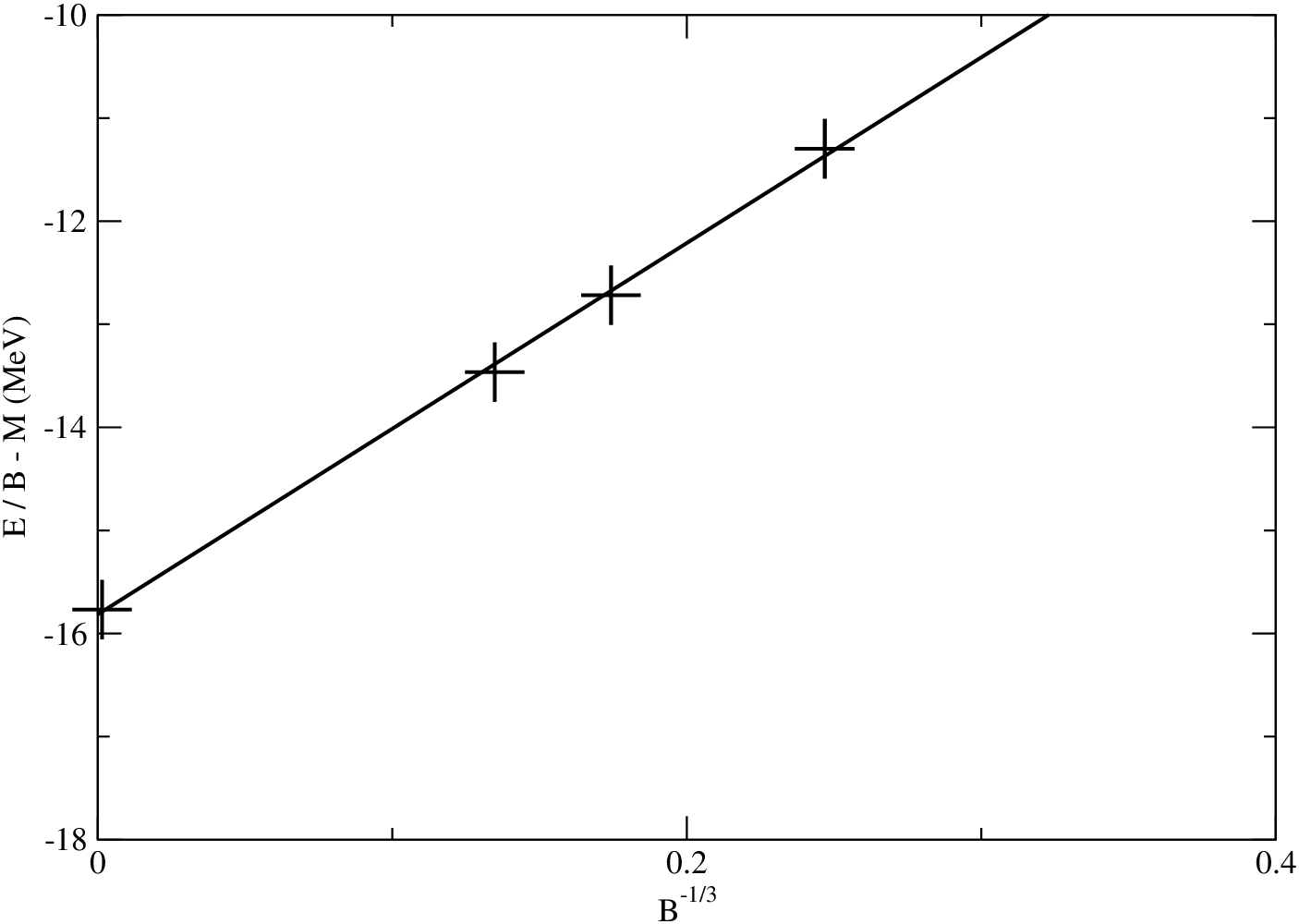}
\caption{Fit to the calculated SEMF for ordinary nuclear matter 
with N = Z and the NLC couplings. The surface energy is 
given by the slope of the curve, here $a_{2} = 18.0$ MeV.}
\label{fig:tf2NLC1SEMFb}
\end{figure}

Now we examine the results of the finite matter investigation.
To begin with, we consider the finite nucleon matter system.
The calculated values of $\mu$, B, $\mathrm{BE = E / B - M}$ as
a function of $r_{0}$ for this type of matter using the L2, NLC, 
and Q1 sets are shown in Table \ref{tab:FNM}. As stated
above, 9 iterations on the vector field were conducted on nuclei 
calculated using the L2 set; this demonstrated the 
convergence of the system. Subsequent finite nucleon matter 
results were obtained using 5 iterations which gave results 
for B and BE to better than $1 \%$. The radii used here, $r_{0} = 15$, 
$20$, and $25$ in units of $m_{s}^{-1}$, include nuclei 
spanning a range of $\mathrm{B} \sim 50 - 400$. As an 
example, $\rho_{B}(r)$ and $\phi_{0}(r)$ for a nucleus with 
$r_{0} = 20m_{s}^{-1}$ calculated with the NLC set 
are displayed in Fig.\ \ref{fig:tf21NLCb}. The interior 
of the nucleus is roughly constant in both 
$\rho_{B}(r)$ and $\phi_{0}(r)$. Then the effective mass
increases to near unity and the baryon density drops
to zero at $r_{0}$; this is a typical example of the surface structure 
for finite nucleon systems.
\begin{table}
\begin{tabular}{|c|c|c|c|c|} \hline
  &$r_{0}$  &$\mu$  &$\mathrm{B}$  &$\mathrm{E / B - M}$  \\ \hline
L2   &15  &4.7055924  &54.568  &-8.7527   \\ \hline 
     &20  &4.69755554 &160.72  &-10.939  \\ \hline
NLC  &15  &4.6968452  &66.438 &-11.297   \\ \hline
     &20  &4.69165108 &188.87 &-12.719  \\ \hline
     &25  &4.68905238 &408.43  &-13.464  \\ \hline
Q1   &15  &4.6965099  &62.581 &-11.262  \\ \hline
     &20  &4.69083149 &179.76  &-12.808  \\ \hline
     &25  &4.68800689 &390.77 &-13.615  \\ \hline
\end{tabular}
\caption{Results of finite nucleon matter for the L2,
NLC, and Q1 parameter sets and various radii. Calculations
with L2 used 9 iterations on the vector field while 5 were
used with NLC and Q1. The radii are in $m_{s}^{-1}$, the 
chemical potential is in $\mathrm{fm^{-1}}$, 
and $\mathrm{E / B - M}$ is in MeV.}
\label{tab:FNM}
\end{table}
Next using the NLC set the calculated values of BE 
are plotted vs.\ $B^{-1/3}$ in Fig.\ \ref{fig:tf2NLC1SEMFb}
for nuclei of various radii. Notice that the infinite
matter value, $\mathrm{BE_{0}}$, has also been included. 
A SEMF of the form Eq.\ 
(\ref{eq:SEMF}) is used as a linear fit in Fig.\ 
\ref{fig:tf2NLC1SEMFb}; the slope of this fit is the surface
energy, in this case $a_{2} = 18.0$ MeV. 
The surface energies for the various coupling sets are given 
in Table \ref{tab:FN} along with the experimentally determined 
value \cite{ref:Wa95}; the values of $a_{2}$ for both NLC and 
Q1 show good agreement with experiment. \emph{The agreement between 
the values calculated with the more realistic interactions
and the empirical result for the surface energy of ordinary 
nucleon matter using this effective lagrangian and 
density functional approach gives us some confidence 
in our exploratory study of the surface
structure of strange superheavy nuclei.}

\begin{table}
\begin{tabular}{|c|c|} \hline
      &$a_{2}$ \\ \hline
L2    &26.510  \\ \hline 
NLC   &18.008  \\ \hline
Q1    &19.107  \\ \hline
Expt. &17.8    \\ \hline
\end{tabular}
\caption{Calculated values of the surface energy (in MeV) for 
nucleon matter using the parameter sets in Table \ref{tab:PAR}.
The experimental value is also included \cite{ref:Wa95}.}
\label{tab:FN}
\end{table}

\begin{figure}
\includegraphics[width=\columnwidth]{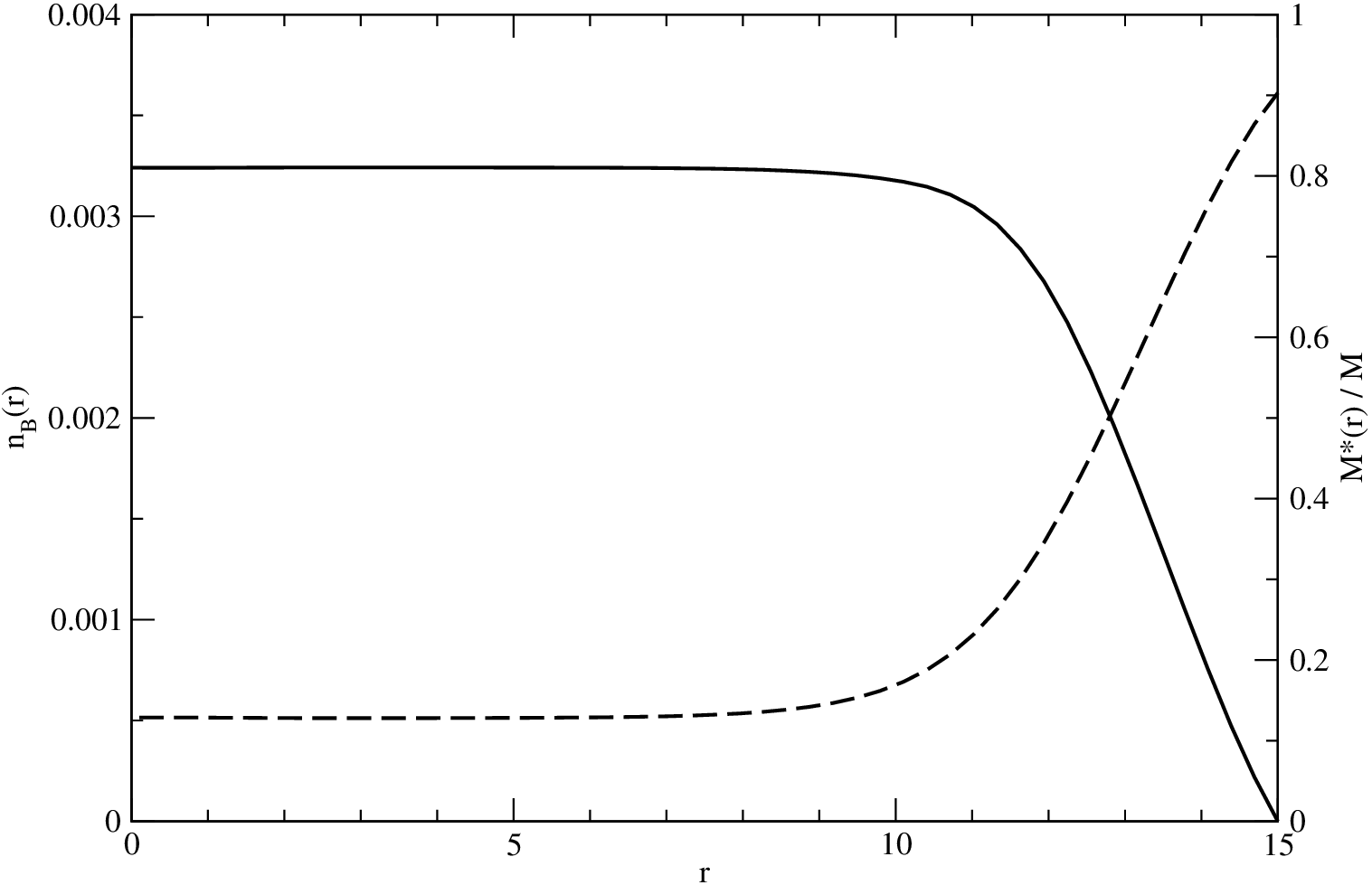}
\caption{Baryon density $n_{B}(r) = \rho_{B}(r) / \mathrm{M^{3}}$ (solid line)
and effective mass $\mathrm{M^{*}(r) / M}$ (dashed line) vs.\ r 
(in units of $m_{s}^{-1}$) for a nucleus composed of nucleons 
and cascades with $r_{0} = 15 / m_{s}$, $\mathrm{B} = 164.918$, and 
$g_{s\Xi} / g_{s} = 1$ subject to the constraints 
$\mathrm{Q} = 0$ and $\mathrm{|S| / B} = 1$. These results were 
obtained using the NLC parameter set.}
\label{fig:tf3NLC15b}
\end{figure}

\begin{figure}[b!]
\includegraphics[width=\columnwidth]{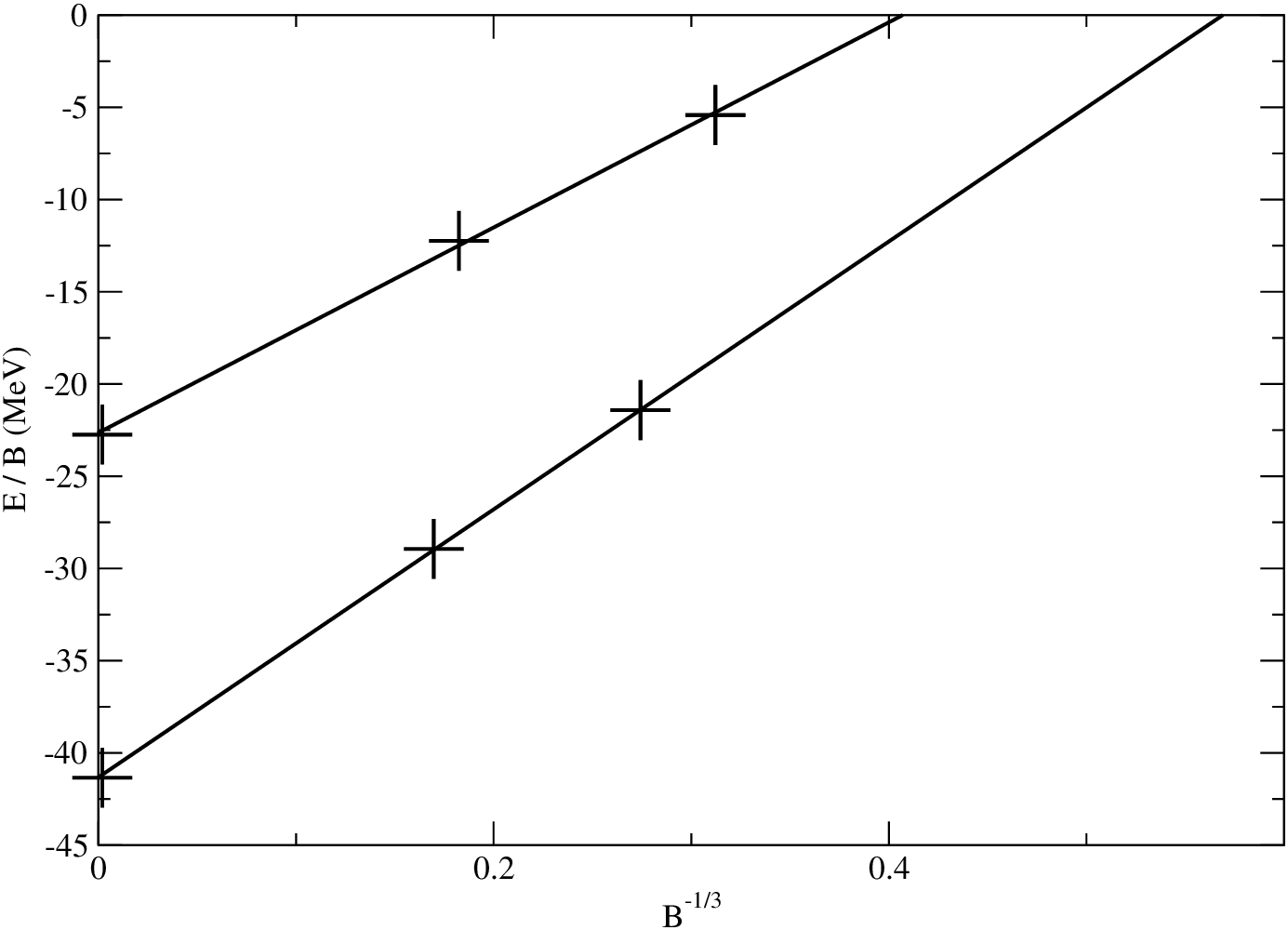}
\caption{Binding energy vs. $\mathrm{B^{-1 / 3}}$ for matter composed of 
equal numbers of cascades and nucleons for the NLC coupling set.
The upper and lower curves correspond to $g_{s\Xi} / g_{s} = .95$ and 1
respectively. The surface energy is just the slope of these lines.}
\label{fig:tf3NLC2SEMFb}
\end{figure}

Now we investigate finite $\Xi N$ matter for the values of 
$g_{s\Xi} / g_{s}$ quoted above. Since the best fit to 
both infinite and finite nucleon matter was obtained with NLC, we 
use this set exclusively in the following discussion. 
The values of $\mu$, B, and $\mathrm{BE} = \mathrm{E / B - M_{\Lambda}}$
obtained for finite $\Xi N$ matter are given 
in Table \ref{tab:FCN}. For the same reasons as in the infinite
$\Xi N$ case, the binding energy per baryon is redefined as
$\mathrm{BE} \equiv \mathrm{E / B - M_{\Lambda}}$. Note that 
due to a slower rate of convergence,
these systems were calculated using 9 iterations on the 
vector field. Also the radii, $r_{0} = 10$ and $15$ in units
of $m_{s}^{-1}$, were
used here; this includes nuclei with baryon numbers ranging 
from $\mathrm{B} \sim 30 - 200$ depending on the $\Xi$ coupling.
It is also important to mention that for the coupling ratio
$g_{s\Xi} / g_{s} = .9$, the nuclei were unbound for 
our choice of radii. For one nucleus of this type, Fig.\
\ref{fig:tf3NLC15b} shows the plots of both $\rho_{B}(r)$ 
and $\phi_{0}(r)$; this is for a nucleus with $g_{s\Xi} / g_{s}
= 1$ and $r_{0} = 15 m_{s}^{-1}$. Notice that the baryon 
densities in the interior of the nucleus
are much larger than those in nucleon matter; also, the effective mass 
drops to less than a third of the nucleon matter
value in the interior. The result is a much higher total B for a 
fixed $r_{0}$. Another feature of note is the surface structure; here
the width of the surface has decreased relative to the previous 
case. For each of the $\Xi$ scalar coupling ratios, the
calculated values of BE are plotted vs.\ $B^{-1/3}$; these
plots are overlaid in Fig.\ \ref{fig:tf3NLC2SEMFb}. The
infinite matter values are also included. As in nucleon matter, 
the SEMF is used as a linear fit, one for each $\Xi$ coupling
ratio, from which the surface energy is determined. The 
values of $a_{2}$ are given in Table \ref{tab:A2CN}. 

\begin{figure}
\includegraphics[width=\columnwidth]{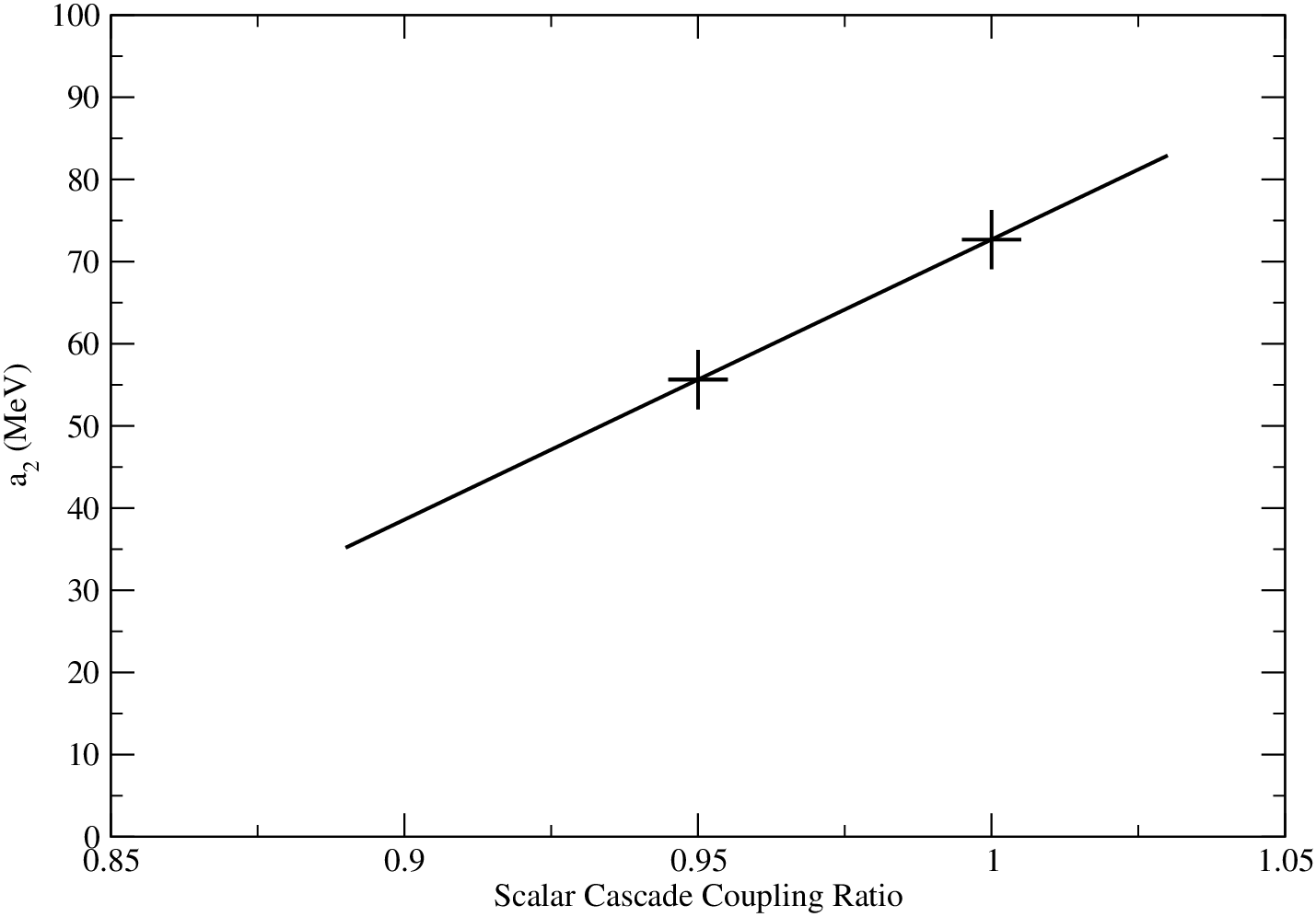}
\caption{Linear fit to the surface energy, $a_{2}$, vs. scalar
cascade coupling ratio ($g_{s\Xi} / g_{s}$) for cascade-nucleon 
matter assuming $\mathrm{Q} = 0$, $\mathrm{|S| / B} = 1$, 
and neglecting $\Lambda$'s.}
\label{fig:tf4NLC2}
\end{figure}

\begin{figure}
\includegraphics[width=\columnwidth]{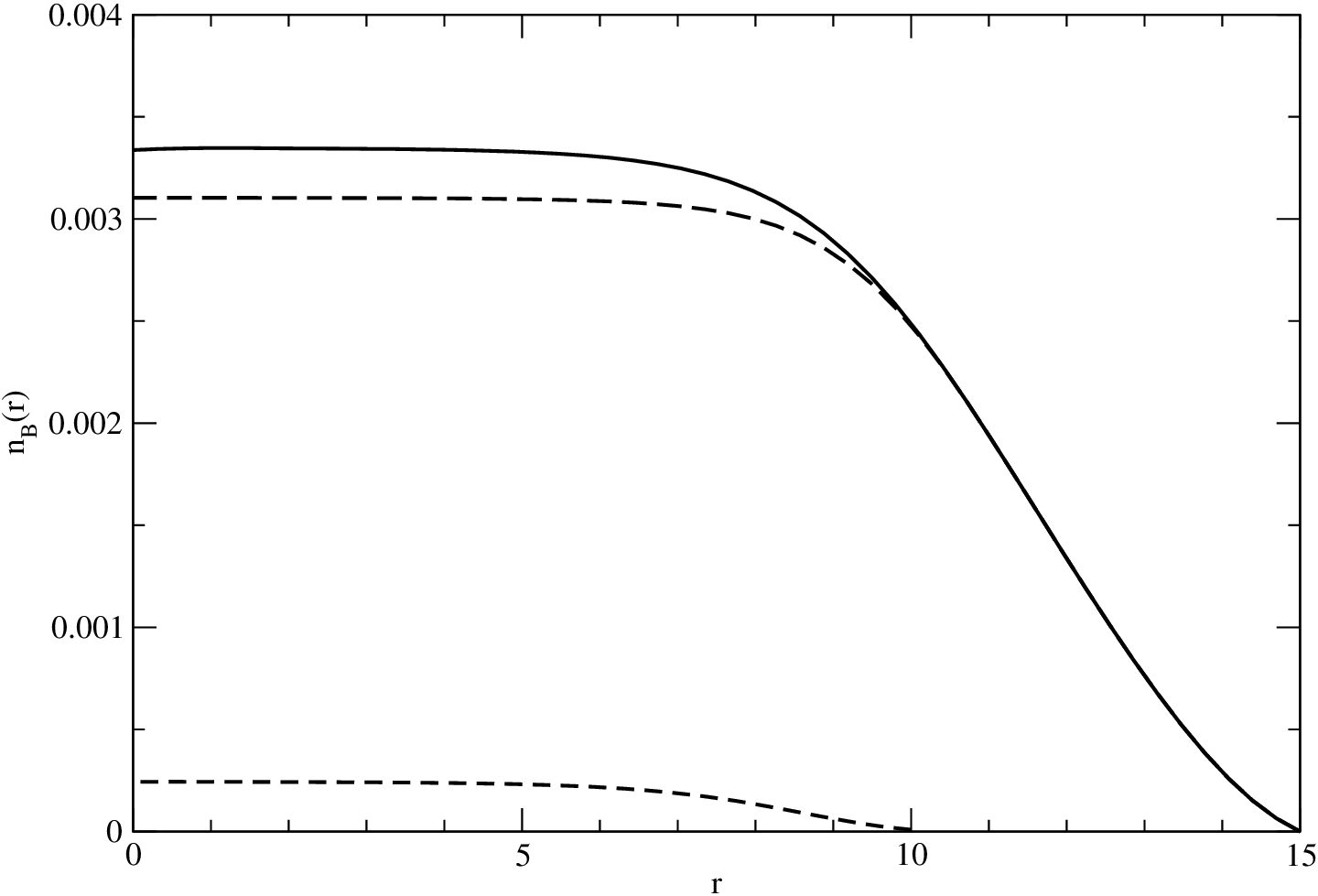}
\caption{Baryon densities for a finite system of nucleons, 
cascades, and lambdas with B = 146.248 for the NLC coupling set and
$g_{s\Xi} / g_{s} = 1$. The total baryon density, the total 
density of cascades and nucleons, and the lambda density are 
shown by the solid, long dashed, and short dashed 
curves respectively. Notice that the lambda density is finite only 
interior to the surface.}
\label{fig:tf4NLC2b}
\end{figure}

As mentioned in section 2, precise data on the binding energy
for a single $\Xi$ in nuclear matter is unavailable. The values
appearing in the literature range 
from -14 to -40 MeV. This necessitated that a number of 
$\Xi$ scalar couplings be investigated. Now that the surface 
energies have been acquired for the $\Xi$ couplings, they 
are plotted vs.\ these coupling ratios in Fig.\
\ref{fig:tf4NLC2}. A linear interpolation is used between
the points; this is extended by extrapolation into
the region which corresponds to values of the binding
energy of a single $\Xi$ appearing in the literature. We
feel confident in the neglect of the $\Lambda$'s over this
region because preliminary 
investigations of finite $\Lambda\Xi N$ matter 
show that the BE and B change little from $\Xi N$ matter. 
In Fig.\ \ref{fig:tf4NLC2b} the baryon density of a 
$\Lambda\Xi N$ nucleus of $r_{0} = 10m_{s}^{-1}$ is shown; 
notice that the $\Lambda$ density begins interior 
to the surface and is comparatively much smaller. 

A preliminary calculation was also conducted with a $\Phi$
meson coupled to the conserved strangeness current. This
simulates a repulsion between like strange particles. In order
to test the size of this interaction which could be tolerated, the 
$\Phi$ coupling was increased until the system was no longer
bound. The values for which this occurred for infinite $\Xi N$ matter are
$g_{\Phi} / g_{\rho} = .68387$, .50902, and .23525 corresponding to 
$g_{s\Xi} / g_{s} = 1$, .95, and .9 respectively. 

It should be mentioned that the conditions $\mathrm{Q} = 0$
and $\mathrm{|S|/B} = 1$ were introduced to eliminate the coulomb
and symmetry terms from the SEMF. Since both charge and 
strangeness are conserved quantities in the strong interaction,
these conditions are unaffected by strong (and electromagnetic) 
reactions in the system.
As a result, this is an intrinsically interesting case and the 
calculation is simplified by the need for only a single Fermi 
wave number. Of course, experimental processes could produce an 
arbitrary Q and $\mathrm{|S|/B}$. Therefore it is of 
interest to estimate how much our results might be modified as these 
conditions are relaxed.

The SEMF has been generalized to include both nucleons and hyperons
in \cite{ref:Do95,ref:Do93,ref:Ba94}. The generalized SEMF proposed 
by Dover and Gal contains additional contributions to the bulk and 
symmetry energies \cite{ref:Do95}. Their SEMF, with their parameter
set I, can be used to estimate how much the quantity $\mathrm{E/B}$
changes as one moves away from the conditions $\mathrm{Q} = 0$ and
$\mathrm{|S|/B} = 1$. The additional terms
result in $\mathrm{|\delta E/B|} < 5$ MeV for the range
$1/2 < \mathrm{|S|/B} < 5/4$ and arbitrary Q. If one makes the 
rough assumption that 
the calculated energy could change by this amount, the surface energy 
extracted from plotting $\mathrm{E/B}$ vs.\ $\mathrm{B}^{-1/3}$ 
could change by up to 30\% (this is undoubtedly an overestimate). 
Calculations with arbitrary Q and $\mathrm{|S|/B}$ 
are more difficult. Work is in progress to examine some of 
these systems.

It is also of some interest to consider the experimental manifestations
of these nuclei. A number of experimental searches for strange matter
have been conducted, examples of which are 
\cite{ref:Hi00,ref:Ba97,ref:Ar01,ref:Ap96}; all have yielded negative 
results. Characteristically, the systems considered here are stable 
against strong decay but unstable against weak decay. Therefore, their 
lifetimes are on the order of the weak interaction timescale, 
or $\sim 10^{-10}$ s. They will experience strangeness changing 
weak decays, such as the decay modes $\Lambda + \mathrm{N} \rightarrow 
\mathrm{N + N}$ and $\Xi + \mathrm{N} \rightarrow \mathrm{\Lambda + N}$. 
We expect heavy ion collisions or supernovae to be possible 
production sources for these systems. Of course, multistrange 
baryon systems would have to make transitions to the ground state 
for the present calculations to be applicable. The actual rate of
production of nuclei of the type considered here is a question that 
goes well beyond the scope of the present paper.

One physical consequence of our results is that the {\it minimum} size of these
objects is predicted. This value is obtained by setting $\mathrm{E/B} 
= 0$ in the SEMF and then solving for B. The minimum baryon 
numbers derived from our calculations are 5.4 and 14.6 for 
$g_{s\Xi}/g_{s} =1$ and .95 respectively. However, shell structure 
becomes more important in the region of small $\mathrm{E/B}$. As a 
result, Hartree calculations are more reliable here. Research is 
in progress to more accurately estimate this number. 

In \emph{conclusion}, we have used this lagrangian density and 
DFT approach to model nucleon, $\Xi N$, and $\Lambda\Xi N$
matter. First, this approach was used to conduct infinite matter
calculations. In the case of infinite nucleon matter, we 
reproduced the equilibrium values from previous works
\cite{ref:Fu97,ref:Se97}. The theory was then extended to
infinite $\Xi N$ and $\Lambda\Xi N$ systems; the equilibrium
values were also found for these types of matter for various
$\Xi$ scalar couplings. The $\mathrm{BE_{0}}$ in each case
was taken to be the bulk term in the SEMF. Then we turned
our attention to finite systems. For the nucleon matter
case, nuclei of various radii were calculated. Plotting
BE vs.\ $\mathrm{B^{-1/3}}$ and using the SEMF as a linear 
fit, the surface energy was extracted. The values for NLC and Q1
were in good agreement with the experimental value \cite{ref:Wa95}.
This gave us the confidence to extend the finite theory
to $\Xi N$ systems. Calculations of $\rho_{B}(r)$ and $\phi_{0}(r)$
gave us a description of the surface structure of these strange 
superheavy nuclei. Again plotting BE vs.\ $\mathrm{B^{-1/3}}$ 
and using the SEMF as a linear fit, the surface energy 
was extracted for a given $\Xi$ coupling ratio. Then using
a linear extrapolation in a plot of $a_{2}$ vs.\ $g_{s\Xi} / g_{s}$,
one can ascertain the surface energy for coupling ratios which
correspond to the single $\Xi$ binding energies in the literature.

I would like to thank Dr. J. D. Walecka for his support,
advice, and for independently verifying much of the numerical
work in this paper. This work was supported in part by DOE grant
DE-FG02-97ER41023. 

\begin{table}
\begin{tabular}{|c|c|c|l|c|c|} \hline
  &$g_{s\Xi} / g_{s}$  &$r_{0}$  &$\;\;\;\;\:\;\;\; \mu$  &$\mathrm{B}$  
&$\mathrm{E / B - M_{\Lambda}}$ \\ \hline
NLC  &1   &10  &1.1604726    &48.459  &-21.416  \\ \hline 
     &    &15  &1.154255303  &204.64  &-28.944  \\ \hline
     &.95 &10  &1.178092     &32.866  &-5.4176  \\ \hline
     &    &15  &1.17206095   &164.92  &-12.241  \\ \hline
\end{tabular}
\caption{Results for finite $\Xi N$ matter for the 
NLC parameter set and a number of radii. These 
calculations used 9 iterations on the vector field. The 
radii are in $m_{s}^{-1}$, the chemical potential is in 
units of $\mathrm{M}$, and $\mathrm{E / B - M_{\Lambda}}$ is in MeV.}
\label{tab:FCN}
\end{table}

\newpage 
\clearpage

\begin{table}
\begin{tabular}{|c|c|c|} \hline
     &$g_{s\Xi} / g_{s}$ &$a_{2}$ \\ \hline 
NLC  &1     &72.687  \\ \hline
     &.95   &55.643  \\ \hline
\end{tabular}
\caption{Values of the surface energy (in MeV) for $\Xi N$ 
matter using the NLC parameter set from Table \ref{tab:PAR}.}
\label{tab:A2CN}
\end{table}

\end{document}